\documentclass[11pt]{article}

\usepackage[margin=1in]{geometry}
\usepackage{graphicx}
\usepackage{authblk}
\usepackage{units}

\title{Using Nab to determine correlations in unpolarized neutron decay}
\author[1]{L.~J.~Broussard} 
\author[2]{S. Bae{\ss}ler} 
\author[3]{T. L. Bailey}
\author[4]{N. Birge}
\author[1]{J. D. Bowman}
\author[5]{C. B. Crawford}
\author[3,6]{C. Cude-Woods}
\author[6]{D. E. Fellers}
\author[4]{N. Fomin}
\author[2]{E. Frle{\v z}} 
\author[7]{M. T. W. Gericke}
\author[8]{L. Hayen}
\author[5]{A. P. Jezghani}
\author[2]{H. Li}
\author[7]{N. Macsai}
\author[6]{M. F. Makela}
\author[9]{R. R. Mammei}
\author[5]{D. Mathews}
\author[6]{P. L. McGaughey}
\author[1]{P. E. Mueller}
\author[2]{D. Po{\v c}ani{\' c}}
\author[3]{C. A. Royse}
\author[2]{A. Salas-Bacci}
\author[6]{S. K. L. Sjue}
\author[1]{J. C. Ramsey}
\author[8]{N. Severijns}
\author[6]{E. C. Smith}
\author[3]{J. Wexler}
\author[4]{R. A. Whitehead}
\author[3]{A. R. Young}
\author[3,6]{B. A. Zeck}

\affil[1]{Oak Ridge National Laboratory, Oak Ridge, TN 37831  USA}
\affil[2]{University of Virginia, Charlottesville, VA 22904, USA}
\affil[3]{North Carolina State University, Raleigh, NC 27695, USA}
\affil[4]{University of Tennessee, Knoxville, TN 37996, USA}
\affil[5]{University of Kentucky, Lexington, KY 40506, USA}
\affil[6]{Los Alamos National Laboratory, Los Alamos, NM 87545, USA}
\affil[7]{University of Manitoba, Winnipeg, MB R3T 2N2, CA}
\affil[8]{Instituut voor Kern-en Stralingsfysica, KU Leuven, Celestijnenlaan 200D, B-3001 Leuven, BE}
\affil[9]{University of Winnipeg, Winnipeg, MB R3B 2E9, CA}

\date{}

\begin{document}

\maketitle

\begin{abstract}
The Nab experiment will measure the ratio of the weak axial-vector and vector coupling constants $\lambda=g_A/g_V$ with precision $\delta\lambda/\lambda\sim3\times10^{-4}$ and search for a Fierz term $b_F$ at a level $\Delta b_F<10^{-3}$. The Nab detection system uses thick, large area, segmented silicon detectors to very precisely determine the decay proton's time of flight and the decay electron's energy in coincidence and reconstruct the correlation between the antineutrino and electron momenta. Excellent understanding of systematic effects affecting timing and energy reconstruction using this detection system are required. To explore these effects, a series of ex situ studies have been undertaken, including a search for a Fierz term at a less sensitive level of $\Delta b_F<10^{-2}$ in the beta decay of $^{45}$Ca using the UCNA spectrometer.
\end{abstract}

\section{Motivational Background} 
\label{intro}

There is strong motivation to use studies of neutron and nuclear beta decay to test our understanding of the electroweak interaction and identify signatures of new physics beyond the Standard Model of Particle Physics. New physics can manifest as a breakdown of unitarity for the Cabbibo Kobayashi Maskawa matrix, with the most precise such test relying on the $V_{ud}$ matrix element from beta decay.  Unitarity tests are particularly sensitive to new physics with vector/axial-vector symmetries, placing broadband constraints at the 11 TeV level for new physics with these symmetries.  Similar constraints for new physics with exotic scalar/tensor interactions can be placed via measurements sensitive to Fierz interference terms. The sensitivity of these low energy probes to new physics is particularly interesting in light of the absence of new particles observed at the Large Hadron Collider. The discovery potential of nuclear and neutron beta decay in comparison to other low and high energy searches was recently reviewed~\cite{Gonzalez-Alonso:2018omy}.

V$_{ud}$ is most precisely determined from the set of superallowed $0^+\rightarrow0^+$ nuclear decays~\cite{Hardy:2014qxa}. The neutron system is an attractive alternative as it is insensitive to nucleus-dependent theoretical corrections. To reach similar precision, the long-standing discrepancy in its measured lifetime must be resolved and its uncertainty improved to $<0.3$\,s, and the ratio of the weak axial-vector and vector coupling constants, $\lambda=g_A/g_V$, must be improved to the $10^{-4}$ level. The value of $\lambda=-1.2724(23)$ recommended by the Particle Data Group~\cite{PDG2018} is defined primarily by measurements of the asymmetry in the decay electron momentum with the neutron polarization, $A_\beta$. $\lambda$ can be determined with independent systematics and similar precision using the correlation between the decay electron and decay antineutrino, $a_{\beta\nu}$.

The Fierz interference term $b_F$ vanishes in the Standard Model but is linear in sensitivity to exotic scalar and tensor interactions. Constraints on scalar and tensor couplings can be constructed from a global analysis of available data from nuclear and neutron beta decay~\cite{Gonzalez-Alonso:2018omy}. The limit $-0.041<b_F<0.225$ (90\% C.L.) has been obtained for the first time in the neutron system from its influence on $A_\beta$ over the measured spectrum~\cite{Hickerson:2017fzz}. There is a compelling case for improving these limits by investigating the impact of $b_F$ on nuclear and neutron beta spectrum shape measurements~\cite{Gonzalez-Alonso:2016jzm,Hayen:2017pwg}. 

\section{The Nab Experiment Overview}
\label{expt}

The primary goal of the Nab collaboration is to perform a determination of $\lambda$ with precision $\delta\lambda/\lambda\sim3\times10^{-4}$. Nab will extract $\lambda$ from the correlation $a_{\beta\nu}$ from the phase space of the proton momentum and electron energy.  The principles of the experiment have been described previously~\cite{Baessler:2012nc}. The 7\,m tall magnetic spectrometer guides charged particles emitted in the neutron decay volume to one of two detectors. Only upward-going protons are permitted to the upper, longer arm of the spectrometer by a narrow 4\,T magnetic field pinch, followed by a magnetic field expansion to more closely relate the proton time of flight to momentum. The spectrometer is currently being commissioned at the Spallation Neutron Source at Oak Ridge National Laboratory and its measured performance meets specifications.

Central to the success of the experiment is the detection system, which must directly detect both the proton and electron in coincidence with excellent timing and energy resolution. The specified performance, in particular 40 ns rise times of the recorded waveforms and 3\,keV full-width half-maximum energy resolution, has been demonstrated using thick, large area, highly segmented silicon detectors and a 24 channel prototype preamplification system, including coincidence detection of neutron decay protons and electrons~\cite{Salas-Bacci:2014xpo,Broussard:2016gqg}. The detectors, electronics, and data acquisition system for the upper detection system in Nab will be floated to a potential of $-30$\,kV to boost the proton's energy so that it can penetrate past the 100\,nm thick dead layer of the silicon. The design of the fully instrumented system has been described~\cite{Broussard:2017tab} and includes improvements over the prototype for more robust coupling to the detectors, better cooling, and improvements to noise performance. The excellent performance of this detection system will also enable a precise determination of the Fierz term with expected precision $\Delta b_F<10^{-3}$. By accepting protons in the lower detection system---instead of the upper detection system---we can improve the counting rate and potentially reduce systematic errors.

\section{Systematics in detection}
\label{sec:syst}

The systematic effects affecting the experiment can be loosely classified: general bias to the data set such as lost events, incorrect assignment of event timing, and incorrect determination of electron energy. Here we focus on the effects of particular import to the detection system: the proton trigger efficiency, the timing profile of charge collection, and the rich challenge of reconstruction of the electron energy measured by the detector. To achieve the uncertainty goals of the experiment, Nab can tolerate an average uncertainty in the efficiency for proton detection of 100 ppm/keV, a bias in the proton time of flight of $\sim$0.3\,ns, and requires a $10^{-4}$ level calibration and understanding of the number of events in the tail of the detected electron energy distribution at the 1$\%$ level.

The proton detection efficiency is reduced by backscattering out of the detector and is especially sensitive to the dead layer of the detector and the noise.  A $-30$\,kV accelerating potential is sufficient to achieve an uncertainty of $<$100 ppm/keV.  The dead layer can vary due to residual gas cryopumping on the detector as observed in~\cite{Broussard:2016gqg} but this effect is negligible in Nab with the high vacuum achieved in spectrometer tests of $5\times10^{-10}$\,Torr. The uniformity of the dead layer and the actual trigger efficiency for protons will be studied using a proton source~\cite{Manitoba}.

The proton time of flight would ideally be determined from the difference in arrival times of the proton and electron at the detector. However, protons and electrons, and electrons of different energies, deposit energy very differently in the detector, resulting in a different timing response and therefore a systematic bias. The uncorrected bias must be below 0.3\,ns to avoid impacting the error budget for $a_{\beta\nu}$. The protons are born with energy $<1$\,keV, and gain 30\,keV from the accelerating potential, and all of their energy is deposited in the first 0.5\,$\mu$m of the 2\,mm thick detector. The electron has up to 782\,keV of energy (minus 30\,keV if impacting the upper detector) and has a distribution of possible deposition profiles up to its maximum range in the detector. 

\begin{figure}[h]
  \includegraphics[width=\textwidth]{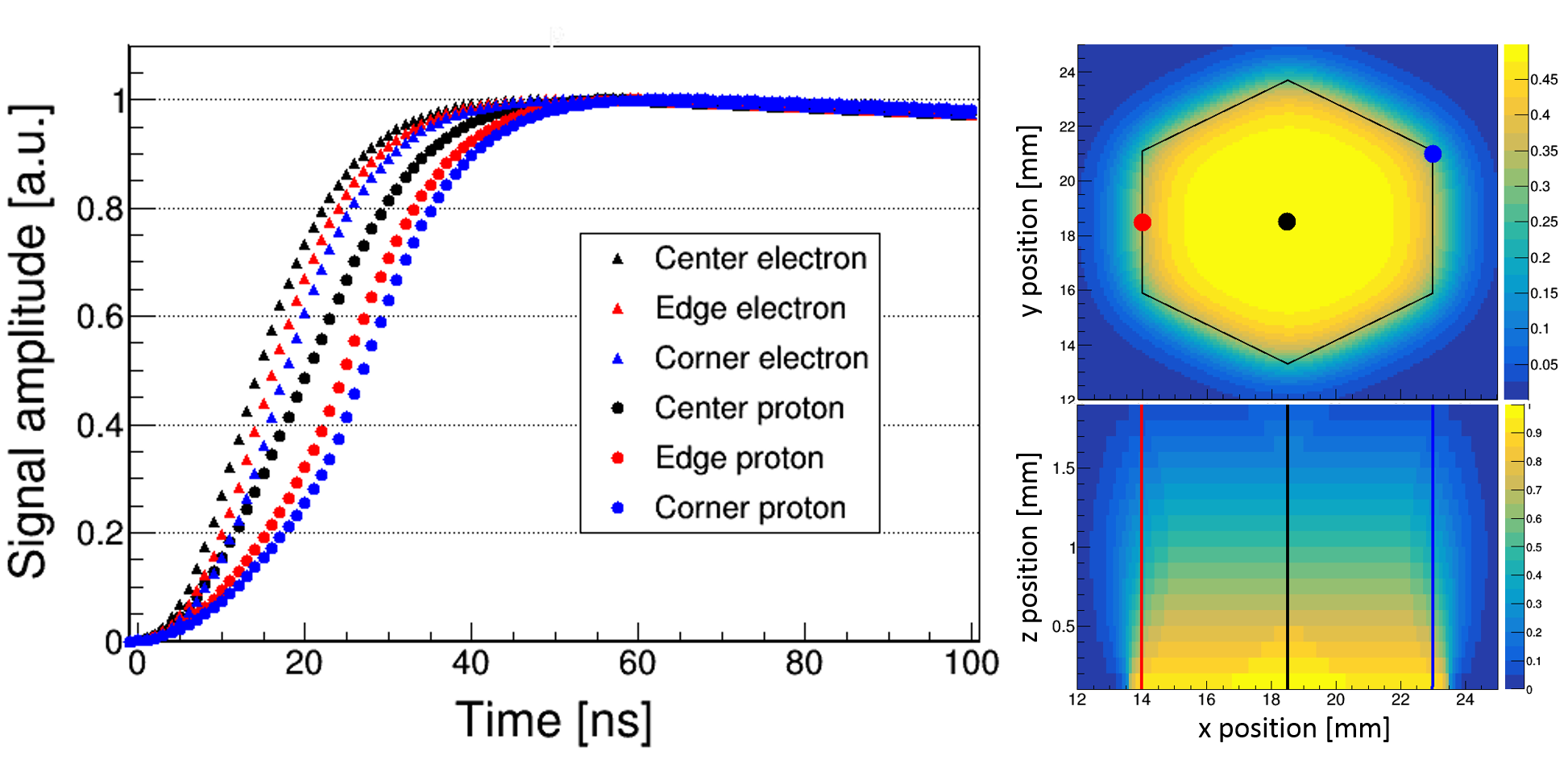}
\caption{Signals induced for electron-hole pairs created at front surface of detector (labeled proton) and middle-depth of detector (labeled electron), at positions in the center of the hexagonal pixel, the edge, and the corner, based on a model assuming uniform impurity density profile. On the right is the weighting potential (top-down and side-view) used to calculate induced signal shapes including drift paths of electron-hole pairs.}
\label{fig:timing}
\end{figure}

The signal induced on a contact from the drift of the electron-hole pairs in the detector can be calculated using the ``weighting field,'' or the component of the electric field from the contact at unit potential and all other contacts grounded, according to the Ramo-Shockley theorem. The signal from quasiparticles created at different locations is shown in Fig.~\ref{fig:timing}, calculated using a modified version of the Majorana \texttt{SIGGEN} code~\cite{siggen}. There is a significant effect on the shape of the rise of the signal from the location of the electron-hole pairs, not only with depth due to the different drift times of the quasiparticles, but also hit position across the surface of the detector due to the variation in weighting potential as induced charge is shared among neighbor contacts. The variation in pulse shape manifests as a bias in the event time extracted from filter or pulse fitting techniques. The average timing shift from the distribution of depths is correlated to the measured energy of the particle. While the hit location is not determined experimentally, the timing could be recovered by summing signals from adjacent pixels (simulating a center hit on a larger pixel) at the cost of increased noise from the increased capacitance.

To characterize our ability to correct for bias caused by charge collection time variations, a fast timing source was developed using a $^{133}$Ba quasi-sealed source closely coupled to a silicon photomultiplier (SiPM) and CeBr3 scintillator assembly. The CeBr3 scintillator crystal and SensL J-series SiPM assembly was selected for its fast timing, good energy resolution, and detection efficiency. The assembly clearly resolved photons with energies as low as 20 keV without cooling or electronics amplification, and was used to detect the fast photons in coincidence with electrons and X-rays detected by the silicon detector. This assembly will be used to build a database of events with known start time, hit position, and energy, to benchmark charge collection simulations and measure the event time bias of various waveform processing algorithms. A variety of conversion electron sources can be used to cover the energy range of interest, or alpha emitters such as $^{241}$Am can be used to compare to protons (range of $\sim$25\,$\mu$m in silicon). In addition to these studies, a low energy electron source with timing information is being developed, and the feasibility of in-situ studies using back-to-back positron annihilation is also being evaluated.

Finally, while the spectrometer is hermetic, the electron loses energy due to various mechanisms before its total energy is collected by one or both detectors. Because one detector has a $-30$\,kV accelerating potential, accurate event-by-event timing reconstruction is needed to understand the bounce history of the electron to determine the energy lost in the inactive dead layers. In addition, as the electron is stopped in the silicon, it loses energy to bremsstrahlung, which has not yet been measured with sufficient precision. Complicating matters, same-pixel hits, as well as accidental coincidences with other decays or backgrounds, result in pile-up. Conversely, multi-pixel hits aid in determining the bounce history, but charge-sharing (physical cross-talk) must be understood to accurately reconstruct the total energy. The number of events in the tail of the detected energy distribution for a given electron energy should be determined with 1\% uncertainty. Finally, the detection system calibration, including linearity, gain, and offset, should be understood to within 0.2\,keV. The detection system will be characterized using a variety of in situ and ex situ studies using sources and pulsers supported by simulations of the geometry using Geant4~\cite{Agostinelli:2002hh} and PENELOPE~\cite{PENELOPE}.

\section{The Fierz term in $^{45}$Ca}   

\begin{figure}[h]
  \includegraphics[width=\textwidth]{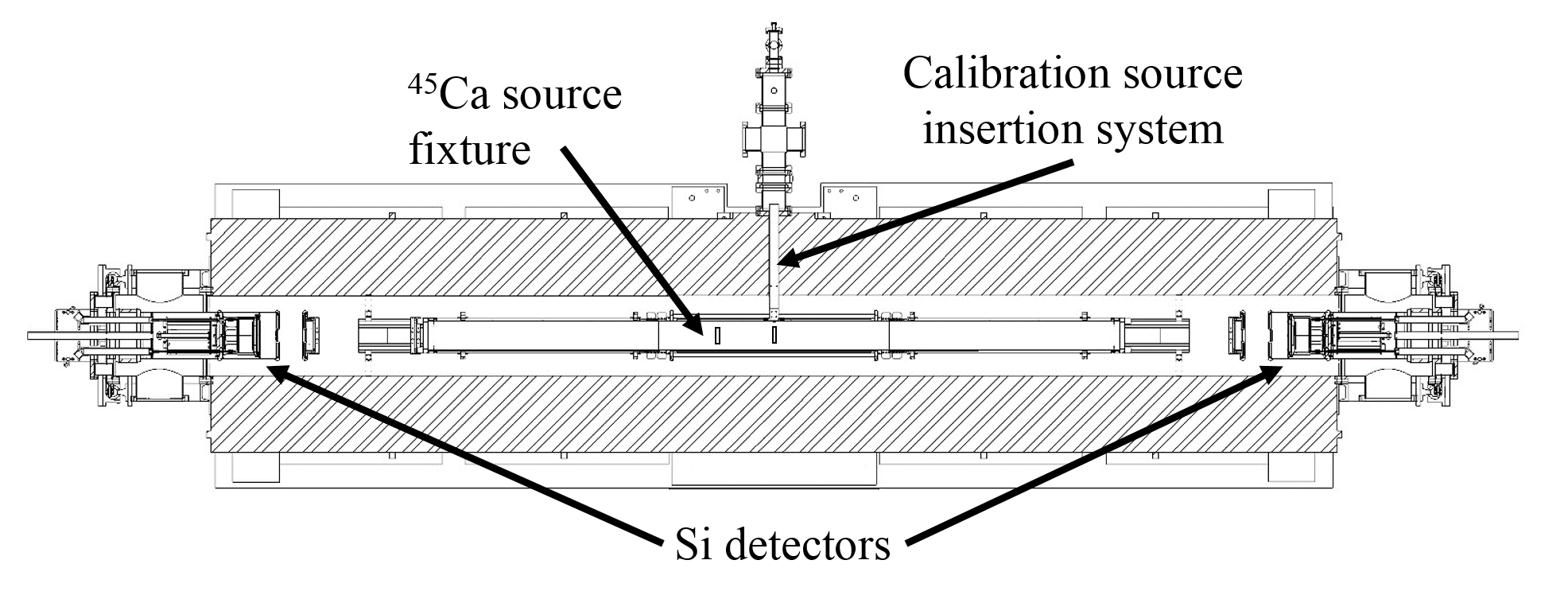}
\caption{Schematic of the experiment to measure the Fierz term in $^{45}$Ca, performed using the UCNA spectrometer with the UCNB detector configuration.}
\label{fig:Ca45}
\end{figure}

To develop our capabilities in precision electron energy reconstruction required for a spectroscopy measurement in Nab, we developed an experiment to measure the Fierz term in $^{45}$Ca using a simpler geometry.  $^{45}$Ca has a kinematic sensitivity to the Fierz term about \nicefrac{2}{3} worse than the neutron due to its lower end-point energy (256\,keV)~\cite{Gonzalez-Alonso:2016jzm}. The experiment was performed using the UCNA spectrometer~\cite{Plaster:2012dma} with the Nab/UCNB prototype detection systems installed at each end (Fig.~\ref{fig:Ca45}). Unlike in Nab, this spectrometer is symmetric with a maximum field of 1\,T at the decay region, the detectors are situated in the expansion region at 0.6\,T, and no accelerating potential is applied to the detectors, however the source foil thickness adds an extra source of energy loss. The source, a deposit of $^{45}$Ca on a 500\,nm foil of Mylar, was installed in the 1\,T region of the spectrometer. To benchmark simulations in Geant4 and PENELOPE, additional studies were performed using conversion electron and X-ray sources ($^{207}$Bi, $^{139}$Ce, and $^{113}$Sn) installed in the spectrometer center (appearing as either in front of or behind the film to either detector) as well as directly in front of each detector. Timing synchronization, gain monitoring, and linearity investigations were available using a pulsed signal applied to the detector bias. Of order $10^{8}$ $^{45}$Ca events were collected, and a total sensitivity of $\Delta b_F<$0.01 is expected.

\section{Summary}   

We have reported on the status of the Nab experiment, which will improve the determination of the parameter $\lambda$ to the $\sim3\times10^{-4}$ uncertainty level and the Beyond Standard Model Fierz term to the $\Delta b_F<10^{-3}$ level. We described the systematic effects most relevant to the silicon detector-based detection system and their impact on the final uncertainty, including the proton trigger efficiency, the charge collection time of protons and electrons, and the electron energy reconstruction.  We have discussed the tools implemented to characterize these effects, including calibration sources, a SiPM+scintillator timing assembly, and proton and electron source facilities.  Finally, we have introduced a new experimental effort to measure the Fierz term in $^{45}$Ca with uncertainty $\Delta b_F<10^{-2}$, currently in analysis, through which the robust energy reconstruction algorithms required for Nab can be developed.

\section*{Acknowledgements}   
Research was sponsored by the Laboratory Directed Research and Development Program [project 8215] of Oak Ridge National Laboratory, managed by UT-Battelle, LLC, for the U. S. Department of Energy, and by the U.S. Department of Energy, Office of Science, Office of Nuclear Physics [contracts {DE-AC05-00OR2272}, {DE-AC52-06NA25396}, {DE-FG02-03ER41258}, {DE-FG02-ER41042}, {DE-SC0008107}, and {DE-SC0014622}], the National Science Foundation [contracts 1126683, 1506320, 1614839, 1615153], the Natural Sciences and Engineering Research Council of Canada [contract {SAPPJ/32-2016}], and the Research Foundation – Flanders (FWO). We are grateful to the UCNA collaboration for use of their spectrometer.

\bibliographystyle{elsarticle-num}
\bibliography{SSP2018arxiv}

\end{document}